\documentclass[aps,dvipsnames]{revtex4-2}
\usepackage[english]{babel}
\usepackage{amsmath,amssymb}
\usepackage{graphicx}
\usepackage{verbatim}
\usepackage{color}
\usepackage{hyperref}
\usepackage{eqnarray,tabularx,picins}
\usepackage{amsthm,esvect,bm,physics,cancel}
\usepackage{wrapfig}
\usepackage{pstricks,pst-circ,pstricks-add,pst-node,pst-coil,pst-eucl,pst-text,pst-fun}

\begin{document}
\title{\color{black}{ Rocket Motion}}
\author{\color{red}{Adel H. Alameh}}
\affiliation{\color{blue}{Lebanese University, Department of Physics, Hadath, Beirut, Lebanon}}
\date{\color{red}{\today}}
\email{adel.alameh@eastwoodcollege.com}

\begin{abstract}
\noindent The motion of rockets is part of the study devoted to the motion of variable mass systems. Notably those in which the mass leaves permanently the considered system. Rockets are propelled forward by the reaction force produced by the hot exhausted gases ejected from their tales in the rearward direction.
Thus their motion should not violate Newton's third principle of the equality of action and reaction forces during the  exhaustion process. Nor should it violate Newton's second law of motion judged by inertial observers. However a close examination of the study of the motion of rockets in a major part of physics textbooks, if not all  reveals erroneous determination of the expression of the thrust force that pushes  the rocket in the forward direction. The false expression of the thrust force entails a bad effect on obtaining the right differential equation that governs the motion of rockets. This trap induced some prominent physics authors to pretend the inapplicability of Newton's second law in such particular cases. Not only that, but they also modified Newton's second law in order to fit their purposes of obtaining the right differential equation, historically known under the name ``Tsiolkovsky rocket equation''.\\
The object of this paper is to give the true expression of the thrust force and to write the differential equation of  motion of rockets without any necessity of modification of the classical laws. The paper also delves into the expression of the change of velocity of rockets,  and proves that their motion is uniformly accelerated in the early stages of liftoff from the ground at condition of constant velocity of expulsion of hot gases from their nozzles.
\end{abstract}
\maketitle
\newpage

\subsection*{Thrust force of a rocket moving in a space free of gravity}

\noindent In this section,  it is intended to study the motion of a rocket moving in deep space far from gravitational effects. At an instant $t$, the mass of the rocket including that of the propellant is $M$ and its velocity is $\bm{V}$ with respect to an inertial observer outside the system. After the passage of certain  time $\Delta t$, the mass of the rocket decreases by $(\Delta M)$ as a result of the fuel consumption and its velocity increases by $\Delta \bm{ V}$. The system (Rocket- ejected fuel) is isolated from all external forces, thus its  linear momentum is conserved. See Fig.1.

\bigskip

\begin{center}
\includegraphics[width=\linewidth]{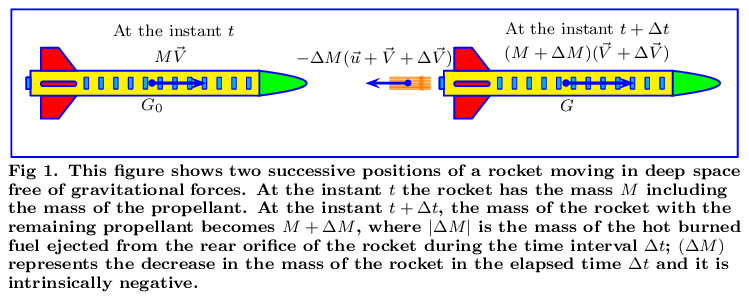}
\end{center}
\bigskip
\noindent Accordingly
\begin{equation}\bm{P}_{initial}=\bm{P}_{final} \label{conservation1}\end{equation}
\begin{equation}M\bm{V}=(M+\Delta M)(\bm{V}+\Delta\bm{V})-\Delta M(\bm{V}+\Delta \bm{V} +\bm{u})\label{conservation2}\end{equation}
where $\bm{u}$ is the velocity of the expelled gases with respect to an observer on the rocket, and $\Delta M<0$.\\

\noindent Simplifying equation (\ref{conservation2}) we obtain$\colon$
\begin{equation}M\Delta \bm{V}=\Delta M \bm{u}\label{conservation3}\end{equation}
dividing equation (\ref{conservation3}) by $\Delta t$ and taking the limit as \mbox{$\Delta t \rightarrow 0$}, we get
\begin{equation}M\displaystyle\frac{d\bm{V}}{dt}=\displaystyle\frac{dM}{dt}\bm{u}\label{newton1}\end{equation}
It is to be noted here that $\displaystyle\frac{d\bm{V}}{dt}$ is the acceleration of the rocket, which has a direction opposite to that of the vector $\bm{u}$ since $\displaystyle\frac{dM}{dt}$ is negative.\\
Unfortunately, the expression $M\displaystyle\frac{d\bm{V}}{dt}$ is incorrectly called the thrust force  exerted  on the body of the rocket by the exhaust gas
~\cite{resnik,wikipedia,mit,terminal,sears}
\begin{equation}\bm{F}_{thrust}=\bm{u}\displaystyle\frac{dM}{dt}\label{falsethrust}\end{equation}
In reality equation (\ref{newton1}) can be written in the form
\begin{equation} \displaystyle\frac{d(M\bm{V})}{dt}-\bm{V}\displaystyle\frac{dM}{dt}=\bm{u}\displaystyle\frac{dM}{dt}\label{thrust1}\end{equation}
which means that
\begin{equation}\displaystyle\frac{d\bm{P}}{dt}=(\bm{u}+\bm{V})\displaystyle\frac{dM}{dt}\label{thrust2}\end{equation}
 and by virtue  of Newton's second law $\bm{F}_{net}=\displaystyle\frac{d\bm{P}}{dt}$, it is obvious that the thrust force on the body of the rocket must be
\begin{equation}\bm{F}_{thrust}=\displaystyle\frac{d\bm{P}}{dt}=(\bm{u}+\bm{V})\displaystyle\frac{dM}{dt}\label{thrust3}\end{equation}
To demonstrate the fallacious conclusion of the expression of the thrust force given in equation (\ref{falsethrust}), we set to find the force exerted by the exhaust fuel on the body of the rocket. To be noted here that the body of the rocket including the remaining unburned fuel is not isolated during the combustion process but it is exposed to the reaction force exerted by the hot expelled fuel. Now according to Newton's second law the net force acting on the rocket is equal to the change of its linear momentum in the interval $\Delta t$. Hence
\begin{equation}\bm{F}_{fuel\rightarrow rocket}\approxeq\displaystyle\frac{(M+\Delta M)(\bm{V}+\Delta \bm{V})-M\bm{V}}{\Delta t}\label{fuelrocket1}\end{equation}
and by taking the limit as $\Delta t \rightarrow 0$ and knowing that  the limit of $\displaystyle\frac{\Delta\bm{V}\Delta M}{\Delta t }$ as $\Delta t \rightarrow 0$ is zero we obtain~\cite{schaum}
\begin{equation}\bm{F}_{fuel\rightarrow rocket}=M\displaystyle\frac{d\bm{V}}{dt}+\bm{V}\displaystyle\frac{dM}{dt}\label{fuelrocket2}\end{equation}
we can replace the expression $M\displaystyle\frac{d\bm{V}}{dt}$ found in equation (\ref{fuelrocket2}) by the expression $\bm{u}\displaystyle\frac{dM}{dt}$ as shown in equation (\ref{newton1})  and the expression of the force exerted by the expelled fuel on the body of the rocket becomes
\begin{equation}\bm{F}_{fuel\rightarrow rocket}=(\bm{u}+\bm{V})\displaystyle\frac{dM}{dt} \label{fuelrocket3}\end{equation}
Now let us calculate the action of the body of the rocket on the hot expelled gases. The force exerted by the body of the rocket on the ejected fuel is equal to the change of the linear momentum of the expelled fuel in the interval $\Delta t$. Therefore
\begin{equation}\bm{F}_{rocket\rightarrow fuel}\approxeq\displaystyle\frac{-(\Delta M)(\bm{u}+\bm{V}+\Delta \bm{V})- (0) \bm{V}}{\Delta t}\label{rocketfuel1}\end{equation}
it is to be noticed here that $\Delta M=0$ at the instant $t$, since at that instant the motor of the rocket starts to burn fuel, and as previously mentioned, the limit of $\displaystyle\frac{\Delta\bm{V}\Delta M}{\Delta t }$ as $\Delta t \rightarrow 0$ is zero, we thus find
\begin{equation}\bm{F}_{rocket\rightarrow fuel}=-(\bm{u}+\bm{V})\displaystyle\frac{dM}{dt}\label{rocketfuel2}\end{equation}
and finally by comparing equations (\ref{fuelrocket3}) with (\ref{rocketfuel2}) we deduce that
\begin{equation}\bm{F}_{rocket\rightarrow fuel}=-\bm{F}_{fuel\rightarrow rocket}\label{actionreaction}\end{equation}
So, the right expression of the thrust force exerted by the exhausted fuel on the body of the rocket is
\begin{equation}\bm{F}_{thrust}=(\bm{u}+\bm{V})\displaystyle\frac{dM}{dt}\label{rightthrust}\end{equation}

\subsection*{Differential equation of motion of a rocket}

\noindent Having established the right expression of the thrust force, we can now seek to write the differential equation that describes the motion of a rocket  when it is projected vertically upward in a uniform gravitational field, neglecting atmospheric resistance. See Fig.2.
Obviously, there are two forces acting on the rocket; its Weight $M\bm{g}$ directed vertically downward and the thrust force $\bm{F}_{thrust}$ in the upward direction. Now we can apply Newton's second law$\colon$
\begin{equation}\Sigma \bm{F}_{ext}=\displaystyle\frac{d\bm{P}}{dt} \label{newton2}\end{equation}
which can be explicitly written as
\begin{equation}M\bm{g}+\bm{F}_{thrust}=\displaystyle\frac{d\bm{P}}{dt} \label{explicit}\end{equation}
and by replacing $\bm{F}_{thrust}$ by its expression given by equation (\ref{rightthrust}) we get
\begin{equation}M\bm{g}+\bm{V}\displaystyle\frac{dM}{dt}+\bm{u}\displaystyle\frac{dM}{dt}=\displaystyle\frac{d{(M\bm{V})}}{dt} \label{thrust1}\end{equation}
which becomes
\begin{equation}M\bm{g}+\bm{V}\displaystyle\frac{dM}{dt}+\bm{u}\displaystyle\frac{dM}{dt}=M\displaystyle\frac{d\bm{V}}{dt}+\bm{V}\displaystyle\frac{dM}{dt} \label{thrust2}\end{equation}
and after canceling $\bm{V}\displaystyle\frac{dM}{dt}$ from both sides of equation (\ref{thrust2}) it becomes
\begin{equation}M\displaystyle\frac{d\bm{V}}{dt}=M\bm{g}+\bm{u}\displaystyle\frac{dM}{dt}\label{final1}\end{equation}
which when projected in the direction of motion gives the differential equation in the algebraic form
\begin{equation}M\displaystyle\frac{dV}{dt}=-Mg-u\displaystyle\frac{dM}{dt}\label{final2}\end{equation}
\begin{center}
\includegraphics[height=8cm,width=5cm]{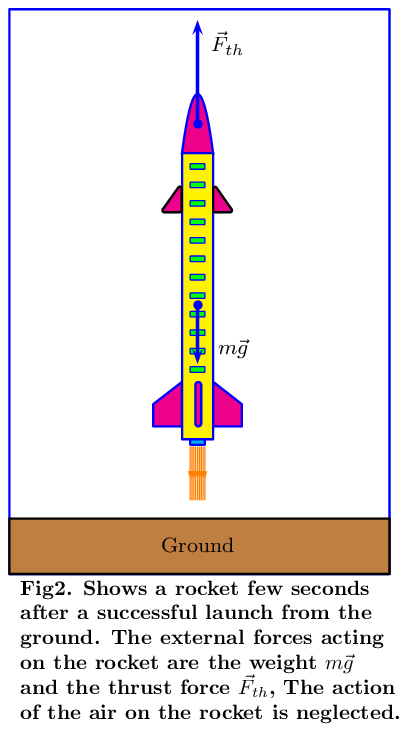}
\end{center}
Although equation (\ref{final2}) is historically well  established but the method of finding it was to apply Newton's second law on the whole system (Rocket + ejected fuel)~\cite{shcaum2}, but when it comes to finding this differential equation by applying the same law on the rocket taken separately, and in view of the erroneous expression of the thrust force; authors were forced to modify the law and put limitations on its validity in order to reach the right equation \cite{plastino,halliday,kleppner}. It is therefore essential to realize that the need for  modification of the classical laws is not justified.
\subsection*{Nature of motion of a rocket in the early stages}
\noindent In this section we shall obtain a solution to the differential equation (\ref{final2}) for a specific particular case where the velocity of expulsion of gases $u$ with respect to an observer on the moving rocket is constant. So we will divide the equation (\ref{final2}) by $M$ and multiply it by $dt$, Hence the equation becomes
\begin{equation}dV=-gdt-u\displaystyle\frac{dM}{M} \label{differential1}\end{equation}
and by integrating the above equation we get
\begin{equation} V=-gt+u\ln(M)+K \label{intergration}\end{equation}
Knowing that the initial velocity of the rocket is \mbox{$V_0=0$} at the instant of liftoff, and that its initial mass is \mbox{$M=M_0$},  we thus obtain the value of the constant $K$ to be \mbox{$K=-\ln(M_0)$}. Substituting the expression of $K$ in the above equation, we obtain
\begin{equation}V=-gt+u\ln\displaystyle\frac{M_0}{M} \label{velocity}\end{equation}
It is to be indicated here that in practice, rockets are preferably designed in a manner as to exhaust fuel at a constant rate. In such cases $\displaystyle\frac{dM}{dt}=-\delta$, where $\delta$ is the rate of consumption of mass per second. $\delta$ is expressed in ($kg/s$). In such a situation the instantaneous mass of the rocket including the remaining fuel at the instant $t$ is given by the linear relation
\begin{equation}M=M_0-\delta t \label{masschange}\end{equation}
$M_0$ being the initial mass of the system \mbox{(rocket+ propellent)}. Thus equation (\ref{velocity}) becomes
\begin{equation}V=-gt+u\ln\displaystyle\frac{M_0}{M_0-\delta t}\label{velocity1} \end{equation}
Here, I suggest to write $\delta$ in the form $\delta =\alpha M_0$ where $\alpha=\displaystyle\frac{\delta}{M_0}$ is a small constant representing the  percentage  in the mass consumption of fuel per second, $0<\alpha<1$ and
$\alpha$ is expressed in $1/s$. On that basis we can write equation (\ref{masschange}) in the form
\begin{equation}M=M_0(1-\alpha t)\label{masschange1}\end{equation}
now by substituting equation (\ref{masschange1}) in equation (\ref{velocity}) we get the expression
\begin{equation}V=-gt+u\ln\displaystyle\frac{1}{1-\alpha t}\label{alpha}\end{equation}
and knowing that the Maclaurin expansion \cite{thomas} of the exponential function has the form
\begin{equation} e^x= 1 +x +\displaystyle\frac{x^2}{2!}+\displaystyle\frac{x^3}{3!}+\ldots \label{maclaurin}\end{equation} and that the expansion given in
equation (\ref{maclaurin}) reduces for small values of $x$, and in a first order approximation to
\begin{equation}e^x=1+x \label{maclaurin1}\end{equation}
One can thus use equation (\ref{maclaurin1}) to say that
\begin{equation} 1-\alpha t=e^{-\alpha t}\label{alphaprime}\end{equation}
And by substitution of equation (\ref{alphaprime}) in equation (\ref{alpha}) one obtains
\begin{equation}V=-gt+u\ln e^{\alpha t}\label{alpha2}\end{equation}
Therefore
\begin{equation}V=-gt +\alpha u t \label{speed}\end{equation}
which becomes
\begin{equation}V=(\alpha u -g)t \label{speed1}\end{equation}
It is evident that as long as $\alpha$ and $u$ remain constant, the expression $(\alpha u -g)$ given in equation (\ref{speed1}) represents the acceleration of the rocket during the early stages of its motion. Furthermore, it does indicate that the velocity of ejection of gases $u$ should satisfy the condition \mbox{$u>\displaystyle\frac{g}{\alpha}$} in order to have a successful liftoff from the ground. Over and above that, having $V$ a linear function of time suggests that the motion of the rocket in the initial phases will be uniformly accelerated.
\subsection*{Conclusion}
\noindent   Having the right expression of the thrust force acting on a rocket enables the applicability of Newton's second law without any need of modification in the case of variable mass systems.

\end{document}